\documentclass[conference]{IEEEtran}
\usepackage{amsmath}
\usepackage{graphicx}
\usepackage{caption2}
\usepackage{amsthm}
\usepackage{cite}
\usepackage{amssymb}
\usepackage{amsfonts}
\begin{document}
\title{DoF Analysis of the K-user MISO Broadcast Channel with Hybrid CSIT}
\author{\IEEEauthorblockN{Borzoo Rassouli\IEEEauthorrefmark{1}, Chenxi Hao\IEEEauthorrefmark{1} and Bruno Clerckx\IEEEauthorrefmark{1}\IEEEauthorrefmark{2}}\\
\IEEEauthorblockA{\IEEEauthorrefmark{1}Communication and Signal Processing Group, Department of Electrical and Electronic Engineering\\Imperial College London, United Kingdom}
\IEEEauthorblockA{\IEEEauthorrefmark{2}School of Electrical Engineering, Korea University, Korea\\
Email: \{b.rassouli12; chenxi.hao10; b.clerckx\}@imperial.ac.uk}
}
\maketitle
\begin{abstract}
%\boldmath
We consider a $K$-user multiple-input single-output (MISO) broadcast channel (BC) where the channel state information (CSI) of user $i(i=1,2,\ldots,K)$ may be either instantaneously perfect (P), delayed (D) or not known (N) at the transmitter with probabilities $\lambda_P^i$, $\lambda_D^i$ and $\lambda_N^i$, respectively. In this setting, according to the three possible CSIT for each user, knowledge of the joint CSIT of the $K$ users could have at most $3^K$ states. Although the results by Tandon et al. show that for the symmetric two user MISO BC (i.e., $\lambda_Q^i=\lambda_Q,\  \forall i\in \{1,2\}, Q\in \{P,D,N\}$), the Degrees of Freedom (DoF) region depends only on the marginal probabilities, we show that this interesting result does not hold in general when $K\geq3$. In other words, the DoF region is a function of all the joint probabilities. In this paper, given the marginal probabilities of CSIT, we derive an outer bound for the DoF region of the $K$-user MISO BC. Subsequently, we investigate the achievability of the outer bound in some scenarios. Finally, we show the dependence of the DoF region on the joint probabilities.\footnote{This work was partially supported by the Seventh Framework Programme
for Research of the European Commission under grant number HARP-318489.}
\end{abstract}

%\begin{IEEEkeywords}
%MISO BC, Alternating CSIT, Degrees of Freedom, Outer Bound, CSIT Pattern
%\end{IEEEkeywords}

\section{Introduction}
In contrast to the point to point multiple-input multiple-output (MIMO) communication where the channel state information at the transmitter (CSIT) does not affect the multiplexing gain, in a multiple-input single-output (MISO) broadcast channel (BC), knowledge of CSIT is crucial for interference mitigation and beamforming purposes \cite{Bruno}. However, the assumption of perfect CSIT may not always be true in practice due to channel estimation error and feedback latency. Therefore, the idea of communication under some sort of imperfection in CSIT has gained more attention recently. The so called MAT algorithm was presented in \cite{MAT} where it was shown that in terms of the degrees of freedom, even an outdated CSIT can result in significant performance improvement in comparison to the case with no CSIT. \cite{Gesbert,Gou12,Chen12a,xinping_Kuser} investigate the time-correlated MISO BC where there is correlation between the feedback information and current channel state, while \cite{Chenxi,Hao} deal with the BC in a frequency-correlated setting.
% Assuming correlation between the feedback information and current channel state (e.g., when the feedback latency is smaller than the coherence time of the channel), the authors in \cite{Gesbert} and \cite{Gou12} consider the degrees of freedom in a time correlated MISO BC which is shown to be a combination of zero forcing beamforming (ZFBF) and MAT algorithm. Following these works, the general case of mixed CSIT and the $K$-user MISO BC with time correlated delayed CSIT are discussed in \cite{Chen12a} and \cite{xinping_Kuser}, respectively. While all these works consider the concept of delayed CSIT in time domain, \cite{Chenxi} and \cite{Hao} deal with the DoF region and its achievable schemes in a frequency correlated MISO BC where there is no delayed CSIT but imperfect CSIT across subbands, which is more inline with practical systems as Long Term Evolution (LTE) \cite{Bruno}.
  In \cite{Tandon} the synergistic benefits of alternating CSIT over fixed CSIT was presented in a two user MISO BC with two transmit antennas. In \cite{Varanasi} and \cite{Amuru}, the MISO BC with hybrid CSIT was considered, however our definition of hybrid CSIT is quite different with that of the aforementioned papers in the sense that instead of having a fixed state (either P or D) for the CSIT of a particular user, the state is allowed to alternate among P, D and N as in \cite{Tandon}.
%The converse in \cite{Tandon} is based on the idea of assigning artificial receivers to the users whose observations are (statistically) equivalent to the corresponding user when CSIT is (not) perfect. However, whether this brilliant approach could be generalized to the scenarios with more than two transmit antennas and two users is unknown. Therefore, for such scenarios, it becomes necessary to check other ways to find the fundamental limits of the system.

Throughout the paper, $f\sim o(\log P)$ is equivalent to $\lim_{P\to\infty}\frac{f}{\log P}=0$ and for a pair of integers $m\leq q$, the discrete interval is defined as $[m:q]=\{m,m+1,\ldots,q\}$. $Y_{[i:j]}=\{Y_i,Y_{i+1},\ldots,Y_j\}$, $Y([i:j])=\{Y(i),Y(i+1),\ldots,Y(j)\}$ and $Y^n=Y([1:n])$.
\section{System Model}\label{s2}
We consider a MISO BC, in which a base station with $M$ antennas sends independent messages $W_1,\ldots,W_K$ to $K$ single-antenna users ($M\geq K$). In a flat fading scenario, the discrete-time baseband received signal of user $k$ at channel use (henceforth, time instant) $t$ can be written as
\begin{equation}\label{equ1}
  Y_k(t)=\mathbf{H}_{k}^{H}(t)\mathbf{X}(t) + W_k(t) \ ,\ k\in[1:K]\ ,\ t\in[1:n]
\end{equation}
where $\mathbf{X}(t)\in \mathbb{C}^M$ is the transmitted signal at time instant $t$ satisfying the (per codeword) power constraint $\sum_{t=1}^n\|\mathbf{x}(t)\|^2\leq nP$. $W_k(t)$ and $\mathbf{H}_{k}(t)(\in\mathbb{C}^M)$ are the additive noise and channel vector of user $k$, respectively, and are also assumed i.i.d. over the time instants and the users.
%Also, let $H(n)={[\textbf{\textit{h}}_1(n),\ldots,\textbf{\textit{h}}_K(n)]}^H$ and $H^n=\{H(1),\ldots,H(n)\}$.
We assume global perfect Channel State Information at Receivers (CSIR) and identity matrix for the covariance of the noise.  %i.e., at time instant $n$, all users have perfect knowledge of $H^n$.

The rate tuple $(R_1,R_2,\ldots,R_K)$, in which $R_i = \frac{\log (|W_i|)}{n}$, is achievable if there exists a coding scheme such that the probability of error in decoding $W_i$ at user $i (i\in[1:K])$ can be made arbitrarily small with sufficiently large coding block length. The DoF region is defined as $\{(d_1,\ldots,d_K)|\exists (R_1,R_2,\ldots,R_K)\in C(P) \mbox{\ such that\ } d_i=\lim_{P\to \infty}\frac{R_i}{\log P},\ \forall i\}$ where $C(P)$ is the capacity region (i.e., the closure of the set of achievable rate tuples).
\begin{figure}[t]
  \centering
  \includegraphics[width=7cm]{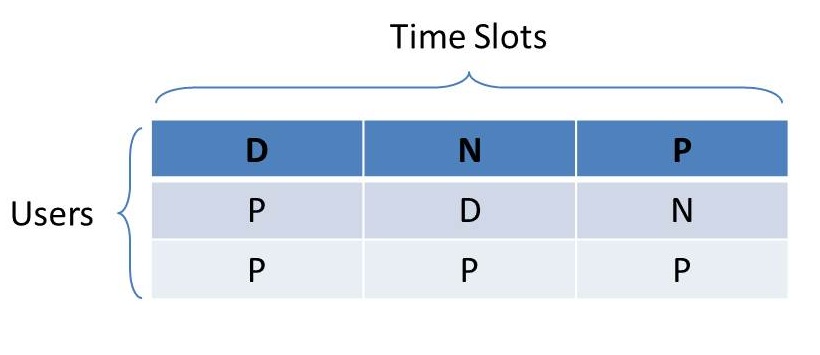}\\
  \caption{A CSIT pattern with $\lambda_{DPP}=\lambda_{NDP}=\lambda_{PNP}=\frac{1}{3}$}\label{fig7}
\end{figure}

The hybrid CSIT model means that at some time instants the transmitter has a Perfect (P) instantaneous knowledge of the CSI of a particular user, whereas at some time instants it receives the CSI with Delay (D) and finally, at some time instants the CSI of the user is Not known (N) at the transmitter. When there is delayed CSIT, we assume that the feedback delay is larger than the coherence time of the channel making the feedback information completely independent of the current channel state. In this configuration, the joint CSIT of all the $K$ users has at most $3^K$ states. For example, in a 3 user MISO BC, they will be $PPP,PDN,\ldots$  with corresponding probabilities $\lambda_{PPP}, \lambda_{PDN},\ldots$\ and the marginal probability of perfect CSIT for user 1 is $\lambda_P^1=\sum_{Q,Q'\in\{P,D,N\}}\lambda_{PQQ'}$.

By CSIT pattern we refer to the knowledge of CSIT represented in a space-time matrix where the rows and columns represent users and time slots, respectively. Figure \ref{fig7} shows an example of a CSIT pattern, in which the transmitter knows the channels of users 2 and 3 perfectly at time slot 1 and has no information about the channel of user 1. The CSI of user 1 will be known in the next time instants (i.e., time slot 2) due to feedback delay and is completely independent of the channel in time slot 2.

The main result of this paper is that given the marginal probabilities of CSIT, an outer bound for the DoF region is provided regardless of the CSIT pattern. Further, through a simple example, we show that the DoF region for $K\geq 3$ must be a function of the CSIT pattern rather than only of the marginal probabilities in contrast to the results of \cite{Tandon} for the 2-user case.  This dependency is equivalent to having the optimal DoF region as a function of $\lambda_{DDP},\lambda_{PNN},\ldots$ in such a way that they do not add up to produce only the marginal probabilities.
\section{Main results}\label{s3}
\textbf{Theorem}. Let $\pi^j(.)$ be an arbitrary permutation of size $j$ over the indices $(1,2,\ldots,K)$, and $\alpha_{\pi^j}(.)$ be a permutation of $\pi^j$ satisfying
\begin{equation}\label{const}
  (\lambda_P^{\alpha_{\pi^j}(i)} + \lambda_D^{\alpha_{\pi^j}(i)})\leq (\lambda_P^{\alpha_{\pi^j}(i+1)} + \lambda_D^{\alpha_{\pi^j}(i+1)})\ \ ,\ \ i\in[1:j-1].
\end{equation}
Given the marginal probabilities of CSIT for user $i$ (which can be any two of $\lambda_P^i, \lambda_D^i$ and $\lambda_N^i$, since $\lambda_P^i + \lambda_D^i + \lambda_N^i = 1$), an outer bound for the DoF region of the $K$-user MISO BC with $M$ transmit antennas at the transmitter ($M\geq K$) is defined by the following sets of inequalities
\begin{align}
 \sum_{i=1}^j\frac{d_{\pi^j(i)}}{i}&\leq 1 + \sum_{i=2}^j\frac{\sum_{r=1}^{i-1}\lambda_P^{\pi^j(r)}}{i(i-1)}\label{theorem1}\\
 \sum_{i=1}^j d_{\pi^j(i)} &\leq 1 + \sum_{i=1}^{j-1}(\lambda_P^{\alpha_{\pi^j}(i)} + \lambda_D^{\alpha_{\pi^j}(i)})
\end{align}
for all $\pi^j$ and $j\in[1:K]$. For the symmetric scenario, the sets of inequalities are simplified as
\begin{align}
  \sum_{i=1}^j\frac{d_{\pi^j(i)}}{i}&\leq 1 + \lambda_P\sum_{i=2}^j\frac{1}{i}
\end{align}
\begin{equation}
    \sum_{i=1}^j d_{\pi^j(i)} \leq 1 + (j - 1)(\lambda_P + \lambda_D).
\end{equation}
For $K=2$, the outer bound boils down to the optimal DoF region in \cite{Tandon}.
\section{Proof of theorem}
%The structure of the proof could be briefly itemized as follows.
%\begin{itemize}
%  \item Applying some sort of improvement to the channel.
%  \item The usage of Fano's inequality.
%  \item Application of the \textit{Csisz\'{a}r sum identity} \cite{network_info} as in \cite{Hao} to change the difference between vector entropies into the sum of the component-wise entropy differences.
%  \item Finding an upper bound for these entropy differences by application of two provided lemmas.
%\end{itemize}
%Having an outer bound for the DoF region of the general $K$-user BC ($M\geq K$),
For simplicity, we assume $j = K$, since it is obvious that each subset of users with cardinality $j$ ($j<K$) can be regarded as a $j$-user BC.
%Therefore, we only consider the proof of the inequalities involving all the $K$ users. For simplicity, we show the inequalities for the
Also, we assume the identity permutation (i.e., $\pi^K(i)=i$) while the results could be easily applied to any other arbitrary permutation.
\subsection{Proof of $\sum_{i=1}^K\frac{d_i}{i}\leq 1 + \sum_{i=2}^K\frac{\sum_{r=1}^{i-1}\lambda_P^r}{i(i-1)}$}
First, we improve the channel by giving the message and observation of user $i$ to users $[i+1:K]$ ($i\in[1:K-1]$). Hence, from Fano's inequality,
\begin{equation}\label{Fano}
  nR_i \leq I(W_i;Y_{[1:i]}^n|W_{[1:i-1]},\Omega^n)+n\epsilon_n
\end{equation}
where $\Omega^n$ denotes the global CSIR up to time instant $n$, $W_0=\emptyset$ and $\epsilon_n$ goes to zero as $n$ goes to infinity.
%This improvement does not decrease the capacity region, meaning that the capacity region of the original channel is a subset of this improved channel. Also,
This improvement results in a degraded broadcast channel \cite{Cover}. Therefore, according to \cite{Elgamal}, since feedback does not increase the capacity of degraded broadcast channels, we can ignore the delayed CSIT (D) and replace them with No CSIT (N).
%In other words, at time instant $n$, knowledge of the CSI up to time instant $n-1$ is not beneficial in a physically degraded BC.
Therefore, it is equivalent to having the channel of user $i$ perfectly known with probability $\lambda_P^i$ and not known otherwise.
%It is important to note that although the channel has become physically degraded, the perfect CSIT (P) cannot be replaced with No CSIT (N), since (P) means that at time instant $n$ the current state of the channel is known to the transmitter perfectly which enables it to know the received signal within noise level (i.e., the results of \cite{Elgamal} cannot be applied in this case.)
From now on, we ignore the term $n\epsilon_n$ for simplicity (since later it will be divided by $n$ and $n\to\infty$) and write
\begin{align}
\sum_{i=1}^K\frac{nR_i}{i}  &\leq  \sum_{i=1}^K\frac{I(W_i;Y_{[1:i]}^n|W_{[1:i-1]},\Omega^n)}{i}  \\
 &\leq h(Y_1^n|\Omega^n) +\sum_{i=2}^K\left[\frac{h(Y_{[1:i]}^n|W_{[1:i-1]},\Omega^n)}{i} \right.\nonumber\\
   &\ \ \left. -\frac{h(Y_{[1:i-1]}^n|W_{[1:i-1]},\Omega^n)}{i-1}\right]+ no(\log P)\label{e1}
\end{align}
where $Y_0=\emptyset$ and we have used the fact that
\begin{equation*}
  \frac{h(Y_{[1:K]}^n|W_{[1:K]},\Omega^n)}{nK}\sim o(\log P).
\end{equation*}
since with the knowledge of $W_{[1:K]}$ and $\Omega^n$, the observations $Y_{[1:K]}^n$ can be reconstructed within the noise distortion.
From the chain rule of entropies, each of the terms in the summation in (\ref{e1}) can be written as
\begin{multline}\label{eq}
%\frac{h(Y_1^n,\ldots,Y_i^n|W_1,\ldots,W_{i-1},H^n)}{i}
%-\frac{h(Y_1^n,\ldots,Y_{i-1}^n|W_1,\ldots,W_{i-1},H^n)}{i-1} \\
\sum_{t=1}^n\left[\frac{h(Y_{[1:i]}(t)|W_{[1:i-1]},Y_{[1:i]}^{t-1},\Omega^t)}{i}\right.
\\ \left.-\frac{h(Y_{[1:i-1]}(t)|W_{[1:i-1]},Y_{[1:i-1]}^{t-1},\Omega^t)}{i-1}\right].
\end{multline}
%where $Y_i^{t-1}$ is the time extension of $Y$ from time instant $i$ to $t-1$.
By adding $Y_i^{t-1}$ in the conditioning of the second entropy, (\ref{eq})  will be increased. Therefore,
%\begin{equation}
%\sum_{t=1}^n\left[\frac{h(Y_{[1:i]}(t)|T_{i,t},\Omega(t))}{i}-\frac{h(Y_{[1:i-1]}(t)|T_{i,t},\Omega(t))}{i-1}\right]
%\end{equation}
\begin{align}
\sum_{i=1}^K\frac{nR_i}{i}&\leq \underbrace{h(Y_1^n|\Omega^n)}_{\leq n\log P}\nonumber\\
&\ \ \ +\sum_{i=2}^K\sum_{t=1}^n\left[\frac{h(Y_{[1:i]}(t)|U_{i,t},\Omega(t))}{i}\right.\nonumber\\&\ \ \ \left.-\frac{h(Y_{[1:i-1]}(t)|U_{i,t},\Omega(t))}{i-1}\right]+ no(\log P)\label{e30}
\end{align}
where $U_{i,t}=(W_{[1:i-1]},Y_{[1:i]}^{t-1},\Omega^{t-1})$ and $\Omega(t)$ is the global CSIR at time instant $t$.
Before going further, the following lemma is needed.

\textbf{Lemma 1}. Let $\Gamma_N=\{Y_1,Y_2,\ldots,Y_N\}$ be a set of $N(\geq2)$ arbitrary random variables and $\Psi_i^{j}(\Gamma_N)$ be a sliding window of size $j$ over $\Gamma_N$ ($1\leq i,j \leq N$) starting from $Y_i$ i.e.,
\[\Psi_i^{j}(\Gamma_N) = Y_{(i-1)_N+1},Y_{(i)_N+1},\ldots,Y_{(i+j-2)_N+1}\]
where $(.)_N$ defines the modulo $N$ operation. Then,
%\begin{multline}\label{e6}
 % (N-m)h(Y_1,Y_2,\ldots,Y_N|A)\leq \sum_{i=1}^{N}h(\Psi_i^{N-m}(\Gamma_N)|A)\nonumber\\
 %  1\leq m\leq N-1
%\end{multline}
\begin{equation}\label{e..6}
  (N-1)h(Y_{[1:N]}|A)\leq \sum_{i=1}^{N}h(\Psi_i^{N-1}(\Gamma_N)|A)
\end{equation}
where $A$ is an arbitrary condition.
\begin{proof} We prove the lemma by induction. It is obvious that (\ref{e..6}) holds for $N=2$. In other words, $h(Y_{[1:2]}|A)\leq \sum_{i=1}^{2}h(Y_i|A)$. Now, considering that (\ref{e..6}) is valid for $N(\geq 2)$, we show that it also holds for $N+1$. Replacing $N$ with $N+1$, we have
\begin{align}
  &Nh(Y_{[1:N+1]}|A)\nonumber\\&=h(Y_{[1:N+1]}|A)+\!(N-1)h(Y_{[1:N-1]},\overbrace{Y_N,Y_{N+1}}^{Z}|A)\nonumber\\
&\leq h(Y_{[1:N+1]}|A)+\sum_{i=1}^{N}h(\Psi_i^{N-1}(\Phi_N)|A)\label{e..8}\\
  &= h(Y_{[1:N+1]}|A)+h(\Psi_1^{N-1}(\Phi_N)|A)\nonumber\\
  &\ \ \ + \sum_{i=2}^{N}h(\Psi_i^{N}(\Gamma_{N+1})|A)\label{e..9}\\
  &= h(Y_N|Y_{N+1},Y_{[1:N-1]},A)\nonumber\\
&\ \ \ +h(\Psi_1^{N-1}(\Phi_N)|A)+h(Y_{N+1},Y_{[1:N-1]}|A)\nonumber\\
&\ \ \ + \sum_{i=2}^{N}h(\Psi_i^{N}(\Gamma_{N+1})|A)\label{e..11}\\
&= h(Y_N|Y_{N+1},Y_{[1:N-1]},A)\nonumber\\
&\ \ \ +h(\Psi_1^{N-1}(\Phi_N)|A)+\sum_{i=2}^{N+1}h(\Psi_i^{N}(\Gamma_{N+1})|A)\nonumber\\
&= h(Y_{N}|Y_{N+1},Y_{[1:N-1]},A)\nonumber
\end{align}
\begin{align}
&\ \ \ +h(Y_{[1:N-1]}|A)+\sum_{i=2}^{N+1}h(\Psi_i^{N}(\Gamma_{N+1})|A)\label{e..13}\\
&\leq h(Y_{N}|Y_{[1:N-1]},A)+h(Y_{[1:N-1]}|A)\nonumber\\&\ \ \ +\sum_{i=2}^{N+1}h(\Psi_i^{N}(\Gamma_{N+1})|A)\label{e13..75}\\
&= h(\Psi_1^{N}(\Gamma_{N+1})|A)+\sum_{i=2}^{N+1}h(\Psi_i^{N}(\Gamma_{N+1})|A) \nonumber\\
&= \sum_{i=1}^{N+1}h(\Psi_i^{N}(\Gamma_{N+1})|A)
\end{align}
%\begin{align}
%  &= \sum_{i=1}^{m}h(\Psi_i^{N+1-m}(\Gamma_{N+1})|A)+\sum_{i=m+1}^{N+1}h(\Psi_i^{N+1-m}(\Gamma_{N+1})|A) \label{e14}\\
%  &= \sum_{i=1}^{N+1}h(\Psi_i^{N+1-m}(\Gamma_{N+1})|A)
%\end{align}
where in (\ref{e..8}), $\Phi_N=\{Y_{[1:N-1]},Z\}$ and we have used the validity of (\ref{e..6}) for $N$. In (\ref{e..9}), we have used the fact that $\Psi_i^{N}(\Gamma_{N+1})=\Psi_i^{N-1}(\Phi_N)$ for $i \in [2:N]$ . In (\ref{e..11}), the chain rule of entropies is used and in (\ref{e..13}), the sliding window is written in terms of its elements. Finally, in (\ref{e13..75}), the fact that conditioning does not increase the differential entropy is used. Therefore, since (\ref{e..6}) is valid for $N=2$ and from its validity for $N(\geq 2)$ we could show it also holds for $N+1$, the proof is complete. \qedhere
\end{proof}
Each term in the summation of (\ref{e30}) can be rewritten as
\begin{equation}\label{e31}
  \frac{(i-1)h(Y_{[1:i]}(t)|U_{i,t},\Omega(t))-ih(Y_{[1:i-1]}(t)|U_{i,t},\Omega(t))}{i(i-1)}
\end{equation}
\begin{align}
  &\leq \frac{\sum_{r=1}^i\left[h(\Psi_r^{i-1}(\Gamma_i)|U_{i,t},\Omega(t))-h(Y_{[1:i-1]}(t)|U_{i,t},\Omega(t))\right]}{i(i-1)} \label{e.3e}\\
  &= \frac{\sum_{r=1}^{i-1}\left[h(Y_i(t)|E_{r,i},U_{i,t},\Omega(t))-h(Y_r(t)|E_{r,i},U_{i,t},\Omega(t))\right]}{i(i-1)}\label{e3e}
\end{align}
where $\Gamma_i=\{Y_{[1:i]}(t)\}$, $E_{r,i} = \{Y_{[1:i-1]}(t)\}-\{Y_r(t)\}$, (\ref{e.3e}) is from the application of lemma 1 and (\ref{e3e}) is from the chain rule of entropies. Before going further, the following lemma is needed.

\textbf{Lemma 2.} In the $K$-user MISO BC defined in (\ref{equ1}), for the users $m,q\in[1:K]$ ($m\neq q$), we have
%\begin{eqnarray}
%% \nonumber to remove numbering (before each equation)
%  Y_m(j) &=& \textit{\textbf{h}}_m(j)^T\textit{\textbf{x}}(j)+w_m(j) \\
%  Y_q(j) &=& \textit{\textbf{h}}_q(j)^T\textit{\textbf{x}}(j)+w_q(j).
%\end{eqnarray}
%Without loss of generality, we assume $m > q$. For simplicity, we assume that the communication is done in real dimensions where $\textit{\textbf{x}}\in R^{M\times 1}$ satisfying $E\left[\|\textit{\textbf{x}}\|^2\right]\leq P$, $\textit{\textbf{h}}_m$ and $\textit{\textbf{h}}_q$ have the distribution $N(\textbf{0},\textbf{I})$ and $w_m$ and $w_q$ have the distribution $N(0,1)$. When the CSIT of a user is either Perfect (P) or Not known (N), the following upper bound holds for the difference between entropies
\begin{equation}\label{e10}
 \lim_{P\to \infty} \frac{h(Y_m(t)|A)-h(Y_q(t)|A)}{\log P}\leq \left\{\begin{array}{cc} 1 & \mbox{CSIT of }{\textbf{H}}_q(t) \mbox{ is } P \\ 0 & \mbox{CSIT of }{\textbf{H}}_q(t) \mbox{ is } N  \end{array}\right.
\end{equation}
where $A$ is a condition such as the condition of entropies in (\ref{e3e}) or later in (\ref{eq1}). Interestingly, (\ref{e10}) is only a function of the CSIT of the second user.
%In other words, in the four possible cases of $PP,PN,NP$ and $NN$, the upper bound (not the exact value) for the pre-log factor of the difference is defined by the CSIT of the second user resulting in the same upper bound for the $PN$ or $NN$ case, and the same upper bound for the $PP$ or $NP$ case.
\begin{proof}
Based on the four possible states for the joint CSIT of ${\textbf{H}}_m(t)$ and ${\textbf{H}}_q(t)$, we have
\subsubsection{CSIT of $\textbf{{H}}_m(t)$ is N or P and CSIT of $\textbf{{H}}_q(t)$ is P}
\begin{equation}\label{e14}
  h(Y_m(t)|A)-h(Y_q(t)|A) \leq \underbrace{h(Y_m(t)|A)}_{\leq \log P}-\underbrace{h(Y_q(t)|A,W_{[1:K]})}_{o(\log P)}
\end{equation}
 A Gaussian input with the conditional covariance matrix of $\Sigma_{X|A}=P\textbf{{u}}_q^{\perp}{\textbf{{u}}_q^{\perp}}^H$ achieves the upper bound, where $\textbf{{u}}_q^{\perp}$ is a unit vector in the direction orthogonal to $\textbf{{H}}_q(t)$ (since $\textbf{{H}}_q(t)$ is known).
\subsubsection{CSIT of $\textbf{{H}}_m(t)$ is N and CSIT of $\textbf{{H}}_q(t)$ is N}
In this case both $Y_m(t)$ and $Y_q(t)$ are statistically equivalent (i.e., having the same probability density functions, and subsequently, the same entropies.) Therefore,
\begin{equation}
  h(Y_m(t)|A)-h(Y_q(t)|A)=0
\end{equation}
\subsubsection{CSIT of $\textbf{{H}}_m(t)$ is P and CSIT of $\textbf{{H}}_q(t)$ is N}
This is shown in \cite{Jafar}.
\qedhere
\end{proof}
From (\ref{e30}) and (\ref{e3e}), we have
\begin{align}
  \sum_{i=1}^K\! \frac{nR_i}{i} &\leq\! \sum_{i=2}^K\sum_{t=1}^n\sum_{r=1}^{i-1}\frac{h(Y_i(t)|A_{r,i,t})-h(Y_r(t)|A_{r,i,t})}{i(i-1)}\\ &\ \ \ +n\log P +no(\log P)\\
%&= n\log P + \sum_{i=2}^K\sum_{r=1}^{i-1}\frac{\sum_{t=1}^n\left[h(Y_i(t)|A(r,i,t))-h(Y_r(t)|A(r,i,t))\right]}{i(i-1)}+no(\log P)\label{e47e} \\
&\leq n\log P + \sum_{i=2}^K\sum_{r=1}^{i-1}\frac{n\lambda_P^r}{i(i-1)}\log P +no(\log P) \label{e4e}
\end{align}
where $A_{r,i,t}$ in the conditioning of the entropies in (\ref{e3e}) and (\ref{e4e}) is from the application of lemma 2 and the fact that $n$ is sufficiently large. Therefore,
\begin{equation}
  \sum_{i=1}^K\frac{d_i}{i}\leq 1 + \sum_{i=2}^K\frac{\sum_{r=1}^{i-1}\lambda_P^r}{i(i-1)}.
\end{equation}
It is obvious that the same approach can be applied to any other permutation on $(1,2,\ldots,K)$ which results in (\ref{theorem1}).
%In addition to the mentioned proof, an alternative converse is provided in Appendix \ref{s4}.
\subsection{Proof of $\sum_{i=1}^K d_i \leq 1 + \sum_{i=1}^{K-1}(\lambda_P^{\alpha_{\pi^K}(i)} + \lambda_D^{\alpha_{\pi^K}(i)})$}
We enhance the channel in two ways:
\begin{enumerate}
  \item Like the approach in \cite{Tandon}, whenever there is delayed CSIT ($D$), we assume that it is perfect instantaneous CSIT ($P$), but we keep the probability of delayed CSIT. In other words, the CSIT of user $i$ is perfect with probability $\lambda_P^i+\lambda_D^i$ and unknown otherwise.
  \item We give the message of user $i$ to users $[i+1:K]$.
\end{enumerate}
  Therefore,
  \begin{equation}\label{e3}
    nR_i\leq I(W_i;Y_i^n|W_{[1:i-1]},\Omega^n)+n\epsilon_n\ ,\ \forall i\in[1:K].
  \end{equation}
   By summing (\ref{e3}) over users and writing the mutual information in terms of differential entropies,
   \begin{align}
     \sum_{i=1}^KnR_i&\leq \overbrace{h(Y_1^n|\Omega^n)}^{\leq n\log P}+ \sum_{i=2}^K\left[h(Y_i^n|W_{[1:i-1]},\Omega^n)\right.\nonumber\\
     &\ \ \ \left.-h(Y_{i-1}^n|W_{[1:i-1]},\Omega^n)\right] + no(\log P).
  \end{align}
By the application of \textit{Csisz\'{a}r sum identity} \cite{network_info}, the term in the summation could be written as
  \begin{equation}\label{e33}
   \sum_{t=1}^n\left[h(Y_i(t)|F_{i,t},\Omega(t))-h(Y_{i-1}(t)|F_{i,t},\Omega(t))\right]
  \end{equation}
  where
  \begin{equation*}
    F_{i,t}=\left (W_{[1:i-1]},\Omega^{t-1},Y_{i-1}^{t-1},Y_i([t+1:n])\right).
  \end{equation*}
Therefore,
 \begin{align}
   \sum_{i=1}^KnR_i&\leq n\log P +\nonumber\\
   &\ \ \sum_{i=2}^K\sum_{t=1}^n\left[h(Y_i(t)|F_{i,t},\Omega(t))-h(Y_{i-1}(t)|F_{i,t},\Omega(t))\right]\label{eq1}
 \end{align}
and finally, by applying the results of lemma 2 to (\ref{eq1}), we have
\begin{equation}\label{e373}
   \sum_{i=1}^K d_i \leq 1+\sum_{i=2}^K(\lambda_P^{i-1}+\lambda_D^{i-1})=1+\sum_{i=1}^{K-1}(\lambda_P^{i}+\lambda_D^{i}).
\end{equation}
Since the same approach holds for any arbitrary permutation of size $K$ on $(1,\ldots,K)$, we have
%\begin{equation}\label{e37}
%   \sum_{i=1}^K d_i \leq 1+\sum_{i=1}^{K-1}(\lambda_P^{\pi^K(i)}+\lambda_D^{\pi^K(i)})\ \ ,\forall \pi^K(.)
%\end{equation}
%(\ref{e37}) results in $K$ inequalities all having the same left hand side. Therefore,
\begin{equation}\label{ee37}
   \sum_{i=1}^K d_i \leq 1+\min_{\pi^K(.)}{\sum_{i=1}^{K-1}(\lambda_P^{\pi^K(i)}+\lambda_D^{\pi^K(i)})}
\end{equation}
and it is obvious that $\alpha_{\pi^K}(.)$ will minimize (\ref{ee37}) if it satisfies (\ref{const}) (for $j=K$.)
\section{Achievability}\label{s55}
In this section, we consider the achievability of the symmetric case.
%The outer bound in theorem consists of $2^K-1+\sum_{j=2}^Kj!\left(\begin{array}{c}K\\j\end{array}\right)$ inequalities. For $K=2$, it is observed that the outer bound (even with $M>2$) will be the same as \cite{Tandon} where it was shown to be achievable, regardless of the pattern of CSIT.
For $K\geq 3$, we show that given the marginal probabilities of CSIT, there exists at least one CSIT pattern that achieves the outer bound in some scenarios. We investigate the following two scenarios:
\subsection{$\lambda_D = 0$}
In this case, $2^K-1$ inequalities are active and the remaining inequalities become inactive
%The reason can be easily verified from the inequalities, however, a simpler intuitive way is to consider that those $\sum_{j=2}^Kj!\left(\begin{array}{c}K\\j\end{array}\right)$ inequalities are derived from making the channel degraded and when there is no delayed CSIT, this degradation results in loose bounds. Equivalently, when there is no delayed CSIT, those inequalities derived from the degraded broadcast channel are inactive.
and the region is defined by $2^K-1$ hyperplanes in $R_+^K$ that has the following K corner points:
\[ (1,\lambda_P,\ldots,\lambda_P),(\lambda_P,1,\lambda_P,\ldots,\lambda_P),\ldots,(\lambda_P,\ldots,\lambda_P,1)\]
%The corner points have the unique characteristic that the whole region can be constructed by time sharing between them. Therefore, the achievability of these points is equivalent to the achievability of the whole region.
%Figure \ref{fig1} shows the region for the 3 user broadcast channel.
The corner points are simply achieved by a scheme that has $N$ time slots and consists of two parts: in the first $\lambda_PN$ time slots, zero forcing beamforming (ZFBF) is carried out where each user receives one interference-free symbol. In the remaining $\lambda_NN$ time slots, only one particular user (depending on the corner point of interest) is scheduled.
%
%\begin{figure}[t]
%  \centering
%  % Requires \usepackage{graphicx}
%  \includegraphics[width=10cm]{dof1}\\
%  \caption{Region in case A for 3 user BC}\label{fig1}
%\end{figure}
\subsection{$\lambda_N\leq \frac{\lambda_D}{\sum_{j=2}^K\frac{1}{j}}$}
Before going further, we need the following simple lemma.

\textbf{Lemma 3}. The minimum probability of delayed CSIT for sending order-$j$ symbols in the $K$-user MAT is
\begin{equation}\label{e9}
  \lambda_D^{min}(K,j)=1-\frac{K-j+1}{K\sum_{i=j}^K\frac{1}{i}}.
\end{equation}
%Substituting $j=1$ in (\ref{e9}), we get the minimum $\lambda_D$ for order-$1$ symbols as
%\begin{equation}
%  \lambda_D^{min}(K)=1-\frac{1}{\sum_{i=j}^K\frac{1}{i}}.
%\end{equation}
%\begin{figure}[t]
%  \centering
%  % Requires \usepackage{graphicx}
%  \includegraphics[width=8cm]{Picture1}\\
%  \caption{Achievable scheme in case A for 3 user BC}\label{fig2}
%\end{figure}
\begin{proof} From \cite{MAT}, the MAT algorithm is based on a concatenation of $K$ phases. Phase $j$ takes $(K-j+1)\binom{K}{j}$ order-$j$ messages as its input, takes $\binom{K}{j}$  time slots and produces $j\binom{K}{j+1}$ order-$j+1$ messages as its output.
%\[(K-j+1)\left(\begin{array}{c}K\\j\end{array}\right) \mbox{order-}j\to \underbrace{{\mbox{Phase }j}}_{\left(\begin{array}{c}K\\j\end{array}\right)\mbox{time slots}}\to j\left(\begin{array}{c}K\\j+1\end{array}\right)\mbox{order-}j+1.\]
In each time slot of phase $j$, the transmitter sends a random linear combination of the $(K-j+1)$ symbols to a subset $S$ of receivers , $|S|=j$. Sending the overheard interferences from the remaining $(K-j)$ receivers to receivers in subset S enables them to successfully decode their $(K-j+1)$ symbols by constructing a set of $(K-j+1)$ linearly independent equations. Therefore, the transmitter needs to know the channel of only $(K-j)$ receivers. In other words, at each time slot of phase $j$, the feedback of $(K-j)$ CSI is enough.
In the MAT algorithm the number of output symbols that phase $j$ produces should match the number of input symbols of phase $j+1$. The ratio between the input of phase $j+1$ and output of phase $j$ is:
\[\frac{(K-j)\binom{K}{j+1}}{j\binom{K}{j+1}}=\frac{(K-j)}{j}.\]
This means that $(K-j)$ repetition of phase $j$ will produce the inputs needed by $j$ repetition of phase $j+1$. In general, in order to have an integer number for repetitions, we multiply phase $1$ by $K!$ (i.e., repeat it $K!$ times), phase $2$ by $\frac{K!}{(K-1)}$, and so on. Therefore, phase $j$ will be repeated $((j-1)!(K-j)!)K$ times which takes $((j-1)!(K-j)!)K\binom{K}{j}$ time slots. Since $(K-j)$ feedbacks from each time slot is sufficient, the number of feedbacks will be $((j-1)!(K-j)!)K\binom{K}{j}(K-j)$. For a successive decoding or order-$j$ symbols, all the higher order symbols must be decoded successfully. Therefore, instead of having delayed CSIT at all time instants from all users, the minimum probability of delayed CSIT is the number of feedbacks from phase $j$ to $K$ divided by the whole number of time slots multiplied by the number of users,
%\[\lambda_D^{min}(K,j)=\frac{\sum_{i=j}^K(i-1)!(K-i)!K\binom{K}{i}(K-i)}
%{\sum_{i=j}^K(i-1)!(K-i)!K\binom{K}{j+1}K}=1-\frac{K-j+1}{K\sum_{i=j}^K\frac{1}{i}}.\] \qedhere
\begin{align*}
    \lambda_D^{min}(K,j)&=\frac{\sum_{i=j}^K(i-1)!(K-i)!K\binom{K}{i}(K-i)}{\sum_{i=j}^K(i-1)!(K-i)!K\binom{K}{j+1}K}\\&=1-\frac{K-j+1}{K\sum_{i=j}^K\frac{1}{i}}.
\end{align*}
\end{proof}
%\begin{figure}[t]
%  \centering
%  % Requires \usepackage{graphicx}
%  \includegraphics[width=10cm]{dof2}\\
%  \caption{Region in case B for 3 user BC}\label{fig3}
%\end{figure}
In this case (i.e., $\lambda_N\leq \frac{\lambda_D}{\sum_{j=2}^K\frac{1}{j}}$),
%the $2^K-K-1$ inequalities having $\sum_i d_i$ (summation with equal weights) in the left-hand side become inactive and the remaining $\sum_{j=1}^Kj!\left(\begin{array}{c}K\\j\end{array}\right)$  inequalities are active which construct $\sum_{j=1}^Kj!\left(\begin{array}{c}K\\j\end{array}\right)$  hyperplanes in $R_+^K$.
The region has $2^K-1$ corner points. In other words, if the coordinates of a point are shown as $(p_1,p_2,\ldots,p_K)$, there are $\binom{K}{j}$ ($j\in[1:K]$) points where $j$ of its $K$ coordinates are $\frac{1+\lambda_P\sum_{i=2}^j\frac{1}{i}}{\sum_{i=1}^j\frac{1}{i}}$ and the remaining $K-j$ coordinates are $\lambda_P$.
%\begin{itemize}
%  \item $\left(\begin{array}{c}K\\1\end{array}\right) \mbox{corner points in the form } (1,\lambda_P,\ldots,\lambda_P),(\lambda_P,1,\lambda_P,\ldots,\lambda_P),\ldots,(\lambda_P,\ldots,\lambda_P,1) $
%  \item $\left(\begin{array}{c}K\\2\end{array}\right) \mbox{corner points in the form } (\frac{2+\lambda_P}{3},\frac{2+\lambda_P}{3},\lambda_P,\ldots,\lambda_P),(\frac{2+\lambda_P}{3},\lambda_P,\frac{2+\lambda_P}{3},\lambda_P,\ldots,\lambda_P),\ldots $
%  \item $\left(\begin{array}{c}K\\3\end{array}\right) \mbox{corner points in the form } (\frac{6+5\lambda_P}{11},\frac{6+5\lambda_P}{11},\frac{6+5\lambda_P}{11},\lambda_P,\ldots,\lambda_P),\ldots $
%  \item \ \ $\ldots, \mbox{and finally,}\left(\begin{array}{c}K\\K\end{array}\right)\mbox{corner points in the form }(\frac{1+\lambda_P\sum_{i=2}^K\frac{1}{i}}{\sum_{i=1}^K\frac{1}{i}},\frac{1+\lambda_P\sum_{i=2}^K\frac{1}{i}}{\sum_{i=1}^K\frac{1}{i}},\ldots,\frac{1+\lambda_P\sum_{i=2}^K\frac{1}{i}}{\sum_{i=1}^K\frac{1}{i}})$
%\end{itemize}
%The region for the 3 user broadcast channel and the achievable scheme are shown in figure \ref{fig3} and figure \ref{fig4}, respectively.
The achievable scheme is based on a concatenation of ZFBF and MAT as follows.
%For the first $K$ corner points listed above, the achievability scheme is the same as that in the previous section (i.e., ZFBF + fixed user scheduling).
For the $\binom{K}{j}$ corner points, we write %in the form $(\frac{1+\lambda_P\sum_{i=2}^j\frac{1}{i}}{\sum_{i=1}^j\frac{1}{i}},\frac{1+\lambda_P\sum_{i=2}^j\frac{1}{i}}{\sum_{i=1}^j\frac{1}{i}},\ldots,\lambda_P,\ldots,\lambda_P)$,
\begin{equation}\
  \lambda_P=\frac{M_1}{N_1}, \lambda_D=\frac{M_2}{N_2}, \lambda_D^{min}(j,1)=\frac{m}{n}
\end{equation}
where $m,n,M_i$  and $N_i$ ($i=1,2$) are integers. Making a common denominator between $\lambda_P$ and $\lambda_D$ we have
\begin{equation}
 \lambda_P=\frac{nM_1N_2}{nN_1N_2}, \lambda_D=\frac{nN_1M_2}{nN_1N_2}.
\end{equation}
%\begin{figure}
%  \centering
%  % Requires \usepackage{graphicx}
%  \includegraphics[width=8cm]{Picture8}\\
%  \caption{Achievable scheme in case B for 3 user BC}\label{fig4}
%\end{figure}
We construct $nN_1N_2$ time slots where the CSIT of each user can be Perfect (P) or Delayed (D) in $nM_1N_2$ or $nN_1M_2$ time slots, respectively. In the first $nM_1N_2$ time slots, ZFBF is carried out. In the remaining $n(N_1N_2-M_1N_2)$ time slots, $j$-user MAT algorithm is done.
At each time slot of the ZFBF part, 1 interference-free symbol is received by each user and in the MAT part, $\frac{n(N_1N_2-M_1N_2)}{1+\frac{1}{2}+\cdots+\frac{1}{j}}$ symbols are sent to each of the users in subset $S$ (with $|S|=j$) where $S$ depends on the corner point of interest. In order to do the MAT algorithm in the second part, the minimum probability of delayed CSIT should be met
\begin{equation}
  nN_1M_2 \geq \lambda_D^{min}(j)n(N_1N_2-M_1N_2)
\end{equation}
Dividing both sides by $nN_1N_2$,
\begin{equation}
  \lambda_D \geq \lambda_D^{min}(j,1)(1-\lambda_P)=\lambda_D^{min}(j,1)(\lambda_D+\lambda_N)
\end{equation}
which results in
\begin{equation}
 \lambda_N \leq \frac{\lambda_D}{\sum_{i=2}^j\frac{1}{i}}.
\end{equation}
Since it should be valid for all $j$, we have
\begin{equation}
 \lambda_N \leq \frac{\lambda_D}{\sum_{i=2}^K\frac{1}{i}}.
\end{equation}
\section{Dependency of the dof region on the csit pattern}\label{sh}
\begin{figure}[t]
  \centering
  \includegraphics[width=8cm]{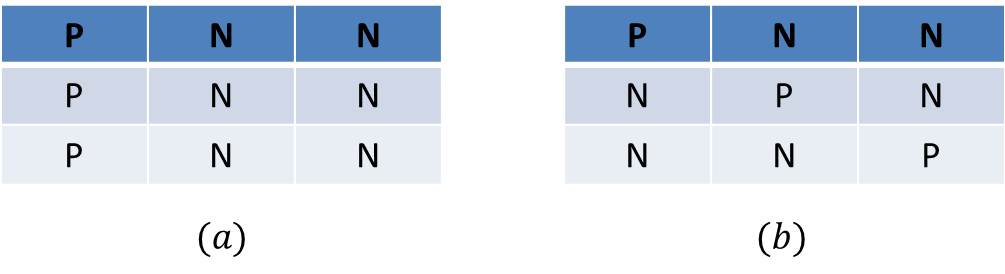}\\
  \caption{Two symmetric CSIT patterns having the same marginal probabilities (i.e., $\lambda_P=1-\lambda_N=\frac{1}{3}$.)}\label{fig100}
\end{figure}

Here, we show that two different CSIT patterns, though having the same marginal probabilities, do not necessarily have the same DoF region. Consider the two simple symmetric CSIT patterns shown in figure \ref{fig100}. According to the theorem, the DoF region of both has an outer bound with the corner points $(1,\frac{1}{3},\frac{1}{3}),(\frac{1}{3},1,\frac{1}{3})$ and $(\frac{1}{3},\frac{1}{3},1)$. It is obvious that the corner points are achievable for pattern $(a)$, and in what follows we show that they are not achievable for pattern $(b)$. We write,
\begin{align}
  nR_1 &\leq I(W_1;Y_1^n|\Omega^n) \label{eq14}\\
  nR_1 &\leq I(W_1;Y_1^n|\Omega^n,W_2)\label{eq15}
\end{align}
Adding (\ref{eq14}) and (\ref{eq15}) results in
\begin{equation}\label{eq16}
  2nR_1 \leq I(W_1;Y_1^n|\Omega^n)+ I(W_1;Y_1^n|\Omega^n,W_2).
\end{equation}
By doing the same for $R_2$, we have
\begin{equation}\label{eq17}
  2nR_2 \leq I(W_2;Y_2^n|\Omega^n)+ I(W_2;Y_2^n|\Omega^n,W_1).
\end{equation}
Finally, the rate of user 3 is written as
\begin{equation}\label{eq18}
  nR_3 \leq I(W_3;Y_3^n|\Omega^n,W_1,W_2).
\end{equation}
Therefore,
\begin{align}
  &2nR_1+2nR_2+nR_3\nonumber\\
  &\leq \underbrace{+h(Y_2^n|\Omega^n,W_1)-h(Y_1^n|\Omega^n,W_1)}_{\leq \frac{n}{3}\log P}+ h(Y_3^n|\Omega^n,W_1,W_2) \nonumber \\
  &\ \ \underbrace{+h(Y_1^n|\Omega^n,W_2)-h(Y_2^n|\Omega^n,W_2)}_{\leq \frac{n}{3}\log P}+\underbrace{h(Y_1^n|\Omega^n)}_{\leq n\log P}+\underbrace{h(Y_2^n|\Omega^n)}_{\leq n\log P}\nonumber\\
  &\ \ \underbrace{-h(Y_1^n|\Omega^n,W_1,W_2)-h(Y_2^n|\Omega^n,W_1,W_2)}_{\leq -h(Y_1^n,Y_2^n|\Omega^n,W_1,W_2)} \nonumber\\&\ \ \ -h(Y_3^n|\Omega^n,W_1,W_2,W_3)\label{53}\\
  &\leq \frac{8n}{3}\log P + h(Y_3^n|\Omega^n,W_1,W_2)-h(Y_1^n,Y_2^n|\Omega^n,W_1,W_2)\nonumber\\
  &= \frac{8n}{3}\log P+\underbrace{h(Y_3^n|T_n)-h(Y_{2,PNN}^n,Y_{1,NPN}^n,Y_{1,NNP}^n|T_n)}_{o(\log P)}\nonumber\\
  &\ \underbrace{-h(Y_{1,PNN}^n,Y_{2,NPN}^n,Y_{2,NNP}^n|T_n,Y_{2,PNN}^n,Y_{1,NPN}^n,Y_{1,NNP}^n)}_{\leq -h(Y_{1,PNN}^n,Y_{2,NPN}^n,Y_{2,NNP}^n|T_n,Y_{2,PNN}^n,Y_{1,NPN}^n,Y_{1,NNP}^n,W_3)\sim o(\log\!P)}\label{a2}\\
  &\leq \frac{8n}{3}\log P
\end{align}
where $T_n = \{\Omega^n,W_1,W_2\}$ and in (\ref{53}), the difference terms are first written as a time summation of instantaneous differences, as in (\ref{e33}). Then, lemma 2 is applied to the differences resulting in the values written under the braces. We have split the observation of users 1 and 2 in terms of the joint CSIT, i.e., $Y_1^n=(Y_{1,PNN}^n,Y_{1,NPN}^n,Y_{1,NNP}^n)$ and $Y_2^n=(Y_{2,PNN}^n,Y_{2,NPN}^n,Y_{2,NNP}^n)$. The first $o(\log P)$ in (\ref{a2}) is due to the fact that there is at least one unknown CSIT (N) in the joint states of user 1 and user 2 (i.e., PN, NP and NN. see rows 1 and 2 of the CSIT pattern shown in figure \ref{fig100} $(b)$) and the fact that (\ref{e10}) is upper bounded by zero when the CSIT of the second term is N.
%Finally, (\ref{a3}) is due to the fact that conditioning reduces the entropy and knowledge of all the messages and the channels enable us to reconstruct each observation within noise distortion.
Therefore, for pattern $(b)$, the following inequalities hold which make its DoF region inside that of pattern $(a)$:
\begin{align}\label{ew1}
  2d_1+2d_2+d_3 &\leq \frac{8}{3} \nonumber\\
  2d_1+d_2+2d_3 &\leq \frac{8}{3} \nonumber\\
  d_1+2d_2+2d_3 &\leq \frac{8}{3}.
\end{align}
Motivated by this simple example, we can have the following set of inequalities for the 3-user MISO BC
%\begin{align}
%  2nR_1+2nR_2+nR_3 &\leq \underbrace{h(Y_1^n|H^n)}_{\leq n\log P}+\underbrace{h(Y_2^n|H^n)}_{\leq n\log P}  \\
%  &\ \underbrace{+h(Y_2^n|H^n,W_1)-h(Y_1^n|H^n,W_1)}_{\leq n(\lambda_P^1+\lambda_D^1)\log P}\underbrace{+h(Y_1^n|H^n,W_2)-h(Y_2^n|H^n,W_2)}_{\leq n(\lambda_P^2+\lambda_D^2)\log P}\\
%  &\ \underbrace{+ h(Y_3^n|H^n,W_1,W_2)\underbrace{-h(Y_1^n|H^n,W_1,W_2)-h(Y_2^n|H^n,W_1,W_2)}_{\leq -h(Y_1^n,Y_2^n|H^n,W_1,W_2)}}_{\leq n(\lambda_{PP-}+\lambda_{PD-}+\lambda_{DP-}+\lambda_{DD-})\log P}
%\end{align}
\begin{align}\label{ew1}
  2d_1+2d_2+d_3 &\leq 2+(\lambda_P^1+\lambda_D^1)+(\lambda_P^2+\lambda_D^2)\nonumber\\&+(\lambda_{PP-}+\lambda_{PD-}+\lambda_{DP-}+\lambda_{DD-})\nonumber \\
  2d_1+d_2+2d_3 &\leq 2+(\lambda_P^1+\lambda_D^1)+(\lambda_P^3+\lambda_D^3)\nonumber\\&+(\lambda_{P-P}+\lambda_{P-D}+\lambda_{D-P}+\lambda_{D-D})\nonumber\\
  d_1+2d_2+2d_3 &\leq 2+(\lambda_P^2+\lambda_D^2)+(\lambda_P^3+\lambda_D^3)\nonumber\\&+(\lambda_{-PP}+\lambda_{-PD}+\lambda_{-DP}+\lambda_{-DD})
\end{align}
where a dash in the above means that the CSIT of the corresponding user is not important (for example, $\lambda_{PD-}=\lambda_{PDP}+\lambda_{PDD}+\lambda_{PDN}$ which is a summation over all the possible values for the CSIT of user 3). The same approach could be easily extended to the $K$-user MISO BC which is omitted for brevity. It is obvious that none of the above inequalities can have its right-hand side written in terms of only marginal probabilities. Therefore, in contrast to the two user scenario, marginal probabilities of CSIT are not sufficient for defining the DoF region of the general $K$-user MISO BC, and having the same marginal probabilities does not guarantee the same DoF region.
\section{Conclusion}\label{s7}
Given the marginal probabilities of CSIT, an outer bound was derived for the DoF region of the $K$-user MISO BC with hybrid CSIT alternating. This outer bound was shown to be achievable by specific CSIT patterns in certain regions. Through an example, we showed that in general, the DoF region of the $K$-user MISO BC (when $K\geq 3$) is a function of CSIT patterns or equivalently the $3^K$ state probabilities rather than the sole marginal probabilities.
\appendices
\bibliography{REFERENCE}

\begin{thebibliography}{10}
\providecommand{\url}[1]{#1}
\csname url@samestyle\endcsname
\providecommand{\newblock}{\relax}
\providecommand{\bibinfo}[2]{#2}
\providecommand{\BIBentrySTDinterwordspacing}{\spaceskip=0pt\relax}
\providecommand{\BIBentryALTinterwordstretchfactor}{4}
\providecommand{\BIBentryALTinterwordspacing}{\spaceskip=\fontdimen2\font plus
\BIBentryALTinterwordstretchfactor\fontdimen3\font minus
  \fontdimen4\font\relax}
\providecommand{\BIBforeignlanguage}[2]{{%
\expandafter\ifx\csname l@#1\endcsname\relax
\typeout{** WARNING: IEEEtran.bst: No hyphenation pattern has been}%
\typeout{** loaded for the language `#1'. Using the pattern for}%
\typeout{** the default language instead.}%
\else
\language=\csname l@#1\endcsname
\fi
#2}}
\providecommand{\BIBdecl}{\relax}
\BIBdecl

\bibitem{Bruno}
B.~Clerckx and C.~Oestges, \emph{MIMO Wireless Networks, 2nd Edition}.\hskip
  1em plus 0.5em minus 0.4em\relax Academic Press, 2013.

\bibitem{MAT}
M.~Maddah-Ali and D.~Tse, ``Completely stale transmitter channel state
  information is still very useful,'' \emph{IEEE Trans. Inf. Theory}, vol.~58,
  no.~7, pp. 4418--4431, 2012.

\bibitem{Gesbert}
S.~Yang, M.~Kobayashi, D.~Gesbert, and X.~Yi, ``Degrees of freedom of time
  correlated miso broadcast channel with delayed {C}{S}{I}{T},'' \emph{IEEE
  Trans. Inf. Theory}, vol.~59, no.~1, pp. 315--328, 2013.

\bibitem{Gou12}
T.~Gou and S.~Jafar, ``Optimal use of current and outdated channel state
  information: Degrees of freedom of the {MISO} {BC} with mixed {CSIT},''
  \emph{IEEE Comms. Letters}, vol.~16, no.~7, pp. 1084 --1087, july 2012.

\bibitem{Chen12a}
J.~Chen and P.~Elia, ``Degrees-of-freedom region of the {M}{I}{S}{O} broadcast
  channel with general mixed-{C}{S}{I}{T},'' vol. arxiv/1205.3474, May, 2012.

\bibitem{xinping_Kuser}
\BIBentryALTinterwordspacing
P.~de~{K}erret, X.~{Y}i, and D.~{G}esbert, ``{O}n the degrees of freedom of the
  {K}-user time correlated broadcast channel with delayed {CSIT},'' in
  \emph{{ISIT} 2013, {IEEE} {I}nternational {S}ymposium on {I}nformation
  {T}heory, {J}uly 7-12, 2013, {I}stanbul, {T}urkey}, {I}stanbul, {TURKEY},
  July 2013. [Online]. Available: \url{http://www.eurecom.fr/publication/4000}
\BIBentrySTDinterwordspacing

\bibitem{Chenxi}
C.~Hao and B.~Clerckx, ``Imperfect and unmatched {CSIT} is still useful for the
  frequency correlated {MISO} broadcast channel,'' in \emph{IEEE ICC},
  Budapest, Hungary, Jun. 2013, available on arXiv:1302.6521.

\bibitem{Hao}
------, ``{MISO} broadcast channel with imperfect and (un)matched {CSIT} in the
  frequency domain: {DoF} region and transmission strategies,'' in \emph{IEEE
  PIMRC}, Sept. 2013.

\bibitem{Tandon}
R.~Tandon, S.~Jafar, S.~Shamai~Shitz, and H.~Poor, ``On the synergistic
  benefits of alternating {CSIT} for the {MISO} broadcast channel,'' \emph{IEEE
  Trans. Inf. Theory.}, vol.~59, no.~7, pp. 4106--4128, 2013.

\bibitem{Varanasi}
K.~Mohanty and M.~Varanasi, ``On the {DoF} region of the {K}-user {MISO}
  broadcast channel with hybrid {CSIT}.'' vol. available on arXiv:1312.1309,,
  Dec. 2013.

\bibitem{Amuru}
S.~Amuru, R.~Tandon, and S.~Shamai, ``On the degrees-of-freedom of the 3-user
  {MISO} broadcast channel with hybrid {CSIT},'' in \emph{IEEE ISIT}, 2014, pp.
  2137--2141.

\bibitem{Cover}
T.~M. Cover and J.~A. Thomas, \emph{''Elements of Information Theory, second
  edition}.\hskip 1em plus 0.5em minus 0.4em\relax New York:
  Wiley-Intersicence, 2006.

\bibitem{Elgamal}
A.~Gamal, ``The feedback capacity of degraded broadcast channels (corresp.),''
  \emph{IEEE Trans. Inf. Theory}, vol.~24, no.~3, pp. 379 -- 381, may 1978.

\bibitem{Jafar}
A.~Davoodi and S.~Jafar, ``Aligned image sets under channel uncertainty:
  Settling a conjecture by {L}apidoth, {S}hamai and {W}igger on the collapse of
  degrees of freedom under finite precision {C}{S}{I}{T},'' \emph{Arxiv
  http://arxiv.org/abs/1403.1541}.

\bibitem{network_info}
A.~E. Gamal and Y.-H. Kim, \emph{Network Information Theory}.\hskip 1em plus
  0.5em minus 0.4em\relax Cambridge University Press, 2012.

\end{thebibliography}
\bibliographystyle{IEEEtran}
\end{document}